\documentclass[amsmath,twocolumn,showpacs,amssymb,aps]{revtex4-1}
\usepackage{graphicx}
\usepackage{geometry}
\usepackage{subcaption}
\usepackage{pgfplots}
\usepackage{dcolumn}
\usepackage{bm}
\usepackage{slashed}
\usepackage{amsmath}
\usepackage{amssymb}
\usepackage{tikz}
\usepackage{bbold}
\usepackage{mathtools}
\usepackage{romannum}
\usepackage{natbib}
\usepackage{braket}

\newcommand{\mP}{\mathcal P}
\newcommand{\mQ}{\mathcal Q}
\begin{document}
\pagenumbering{arabic}
\title{Complexity in the Lipkin-Meshkov-Glick Model \\}
\author{Kunal Pal}
\email{kunalpal@iitk.ac.in}	
\author{Kuntal Pal}
\email{kuntal@iitk.ac.in}
\author{Tapobrata Sarkar}
\email{tapo@iitk.ac.in}
\affiliation{
Department of Physics, Indian Institute of Technology Kanpur, Kanpur 208016, India}
\date{\today}
\begin{abstract}
We study complexity in a spin system with infinite range interaction, via the paradigmatic 
Lipkin-Meshkov-Glick model, in the thermodynamic limit. Exact expressions for the Nielsen complexity 
(NC) and the Fubini-Study complexity (FSC) are derived, that helps us to establish several distinguishing features compared
to complexity in other known spin models. 
In a time-independent LMG model, close to phase transition, the NC diverges logarithmically, much like the entanglement 
entropy. Remarkably however, in a time dependent scenario, this divergence is replaced by a finite discontinuity, as we show by
using the Lewis-Riesenfeld theory of time-dependent invariant operators. The FSC of a variant of the LMG model 
shows novel behaviour compared to quasi free spin models. 
Namely, it diverges logarithmically when the target (or reference) state is near the separatrix. 
Numerical analysis indicates that this is due to the fact that geodesics starting with arbitrary boundary conditions
are ``attracted'' towards the separatrix and that near this line, a finite change in the affine parameter of the geodesic produces an
infinitesimal change of the geodesic length. The same divergence is shared by the NC of this model as well. 
\end{abstract}
\maketitle

\section{Introduction}
\label{intro}

The application of information theoretic methods to study the physics of zero temperature quantum phase transitions \cite{Sachdev}
has a long history (for a recent review, see \cite{Carollo}, see \cite{Gu} for an older one). 
In the last few years, a new information theoretic measure, the complexity,
has been popular in the literature. Originally introduced in computer science
and related studies quite a while back, the notion of complexity has gained importance after it was realised that this
might offer deep insights into the physics of black holes \cite{Susskind1}-\cite{Susskind4}, via the gauge-gravity 
duality of high energy physics.

There are broadly two popular notions of complexity at present. The first was introduced by Nielsen in the 
context of quantum computation \cite{Nielsen1}-\cite{Nielsen3}. This version of the ``circuit complexity,'' also  
dubbed as the Nielsen complexity (NC) is a measure of  the minimum number of unitary gates required to make a desired operator,
starting from a reference operator. 
This is obtained by considering a trajectory in the space of unitary transformations that minimises the cost 
(or length) functional in such a space. Among several possible variants of such a functional, 
in this paper, we will consider only the so called ``$\mathcal{F}_{2}$ cost functional" \cite{Myers1}.
The NC has been widely studied in high energy physics of late \cite{Myers1}-\cite{ABHKM}.
Besides, spin chain models of statistical mechanics also provide a convenient backdrop for studying NC,
and such studies have been carried out in \cite{LWCLTYGG}-\cite{WH}.
 While the NC is related to geodesic distance on the space of unitary operators,  
another measure of the complexity of a quantum mechanical system was introduced in \cite{CHMP}, 
namely the geodesics distance between two points on the Riemannian manifold of the parameter space of the system. 
This is often equipped with a Fubini-study metric \cite{PV}, and thus this measure of complexity 
is called the Fubini-Study complexity (FSC). This was studied in details the context of QPT in the transverse field 
Ising model in \cite{JGS1}, \cite{JGS2}, where the relation between the NC and the FSC was established for quasi free
fermionic systems. 
 
In the context of information geometry, many studies focus on the so called XY model \cite{zanar} and its variations. 
This is mainly due to the fact that this model, which incorporates nearest-neighbour interactions only, 
can be mapped to a quasi-free fermionic system which makes the analysis simpler, and offers an explicit
relation between the NC and the FSC in certain limits \cite{JGS1}, \cite{JGS2}. A significantly different but
non-trivial example is the Lipkin-Meshkov-Glick (LMG) model of nuclear physics \cite{LMG}. The LMG model incorporates
long range interactions which induces a quantum phase transition in the thermodynamic limit \cite{Botet1},\cite{Botet2}. 
In the context of information geometry, the LMG model is a curious example, 
as the QMT is a-priori ill defined in this case \cite{SMK, dey}. As a result, 
the standard way of exploring the zero-temperature phase transition is not possible here. 
A significant step in this direction was taken in \cite{GGCHV}, where a suitable modification to 
the classical LMG Hamiltonian was shown to cure this problem and it was shown there that the QMT 
is well defined, with a compact expression. 

In this backdrop, an important and interesting question
is the behaviour of the FSC. As we show in this paper, the FSC here shows novel features that have not been 
seen in other spin models studied till date. Namely, geodesics are ``attracted" towards the separatrix of
the model and as a result the geodesic length diverges near the QPT, as finite increments in an affine parameter
that parametrises the geodesic only results in infinitesimal changes in the geodesic length near this line. 
This is in stark contrast to the XY model, where the the derivative of the FSC shows 
logarithmic divergence near a phase transition (a fact that can be gleaned from the scaling
of the information metric), with the FSC itself being regular there.   

On the other hand, in the context of the NC, we note that in the thermodynamic limit, the ground state of the  
LMG model is well approximated by a Gaussian wavefunction. 
For such wavefunctions, the NC has been computed recently both in time independent \cite{Myers1} and time 
dependent \cite{ABHKM} cases -- results that have found wide usage in quantum field theory of late. 
One can then envisage using these results for computing the NC in the LMG model. 
We show in this paper that this is indeed possible, and shows interesting results. Namely, while in the time-independent
case the NC diverges due to the presence of the ``zero mode,'' this is replaced by a finite discontinuity 
in the time-dependent case, as we establish via the Lewis-Riesenfeld formalism. 
Clearly then, the above discussions indicate that one can compute and compare the 
different measures of complexity in the LMG model. This is the purpose of this paper, and the details are 
elaborated upon in the next sections. This paper contains five appendices, to which we relegate some of the 
mathematical details. Throughout the rest of this paper, we will set $\hbar = 1$. 

\section{The  LMG model and Nielsen complexity of the  ground state}
\label{sec1}

The Hamiltonian of the Lipkin-Meshkov-Glick model of nuclear physics  
describes a collection of $N$ self interacting spin $1/2$ particles acted upon by an external field, and is given by
\begin{equation}
\label{LMGH}
H =-\frac{2}{N}\left(J^{2}_{x}+
\gamma J^{2}_{y} \right)-2 BJ_{z}+\frac{\kappa}{2}\left(1+\gamma\right)~.
\end{equation}
Here $B$ is an external  magnetic field which is assumed to be along the $z$ direction, and $J_{\alpha}$, 
with $\alpha=\{x,y,z\}$, are the components of the collective spin 
operator $J$. These components can be written in terms of the spin operator of the 
individual particle as $J_{\alpha}=\sum_{i}\sigma^{i}_{\alpha}/2$. The constant 
$\gamma$  characterise the anisotropy of the spin-spin interaction, and the last term of the 
Hamiltonian is a constant energy shift \cite{CCPPF} and depends on the constant $\kappa$. 
In this paper, we assume $\kappa>0 $ and $0\leq\gamma\leq1$. 

Phase transitions in the LMG model have been extensively studied (see for example \cite{RVM1,RVM2}) and as is 
well known, in the thermodynamic limit $N \rightarrow \infty$, the ground state of the LMG model shows a second 
order phase transition for the critical value of 
the magnetic field $B_{c}=1$. Our goal in this section is to calculate the Nielsen circuit  
complexity between ground states for two values of the parameter of the LMG model, in this limit. 
Using the standard Holstein-Primakoff (HP) transformation in this limit, the Hamiltonian of  
Eq. (\ref{LMGH}) can be mapped to that of a harmonic oscillator \cite{DV1,DV2}. 
The details of this  mapping can be found in \cite{CCPPF,DV2}, and we write down here the final form of the 
Hamiltonian after the HPT and a subsequent Bogoliubov transformation by which the Hamiltonian is reduced to a 
convenient diagonal form written in terms of a new set of creation and annihilation operators.

Assuming $B\geq0$ we need to consider two different phases before and after the quantum phase transition (QPT).\\
\noindent
{\bf Case 1 - Symmetry broken phase (BP): $0\leq B \leq 1$.} Here, the final form of the Hamiltonian, 
written after the above mentioned transformations, in terms of a new set of bosonic operators  
$\{c^{\dagger},c\}$ is given by
\begin{eqnarray}\label{LMG2}
\begin{split}
H_{BP}=&2\sqrt{\left(1-B^{2}\right)\left(1-\gamma\right)}
\left(c^{\dagger}c+\frac{1}{2}\right)\\
&-\frac{1+B^{2}N}{2}-\frac{1-\gamma}{2}~.
\end{split}
\end{eqnarray} 
These operators are related to original set of Bosonic operators used in the HPT, called $\{a^{\dagger},a\}$, 
through the Bogoliubov transformation. Since $\{a^{\dagger},a\}$ are related to collective spin operators, 
$\{c^{\dagger},c\}$ can be written in terms of $J_\alpha$. Again, the details of these transformations, 
which can be found in \cite{DV2,CCPPF}, are omitted here. \\
\noindent
{\bf Case 2 - Symmetric phase (SP): $B > 1$.} In this case, the final Hamiltonian after the HPT and the 
Bogoliubov transformations is given by 
\begin{eqnarray}\label{LMG1}
\begin{split}
H_{SP}=&2\sqrt{\left(B-1\right)\left(B-\gamma\right)}\left(b^{\dagger}b+\frac{1}{2}\right)\\
&-B\left(N+1\right)+\frac{1+\gamma}{2}~.
\end{split}
\end{eqnarray}
In both the phases the final form of the Hamiltonian represents a harmonic oscillator with frequency 
$\omega_{1}=2\sqrt{\big(1-B^{2}\big)\big(1-\gamma\big)}$ in the BP and    
$\omega_{2}=2\sqrt{\big(B-1\big)\big(B-\gamma\big)}$ in the SP. In the limit $B \rightarrow 1$ in both the phases, 
the frequency goes to zero so that energy gap between different levels vanishes and the system becomes infinitely degenerate.

We want to calculate the Nielsen complexity between two ground states both of which are in either phase of the system 
(i.e., both the states are either in the BP or the SP) with different values of the parameters. For this 
we need the expressions for the wavefunctions in both the phases. 
It is convenient to define the position and the momentum operators in  the BP and SP respectively  as follows
\begin{eqnarray}\label{coordinatescons}
Q_{1}=\frac{1}{\sqrt{2\omega_{1}}}\Big(c^{\dagger}+c\Big)~,~P_{1}=i\sqrt{\frac{\omega_{1}}{2}}\Big(c^{\dagger}-c\Big)\nonumber\\
Q_{2}=\frac{1}{\sqrt{2\omega_{2}}}\Big(b^{\dagger}+b\Big)~,~P_{2}=i\sqrt{\frac{\omega_{2}}{2}}\Big(b^{\dagger}-b\Big)~.
\end{eqnarray} 
Here, the subscript $1$ and $2$ respectively indicates the BP (with $0\leq B \leq 1$) and the SP ($B>1$).
The same notations are used throughout the text. 
In the position space then the ground state wavefunctions of the systems corresponding to the two 
phases are just Gaussians of the form
\begin{equation}
\Psi(Q_i)=\mathcal{N}\exp\Big[-\frac{1}{2}\omega_iQ_i^2\Big]~.
\end{equation} 
Now, using the standard formula for the Nielsen complexity of preparing such a Gaussian wavefunction from another one 
of different frequency \cite{Myers1}, we have, in the BP of the system, the complexity of creating a 
state with frequency $\omega_{1T}$ starting from a reference state with frequency $\omega_{1R}$ to be given by 
\begin{equation}
\mathcal{C}_{N}^{BP}=\frac{1}{2}\ln \Bigg[\frac{\omega_{1T}}{\omega_{1R}}\Bigg]=\frac{1}{4}\ln \Bigg[
\frac{\big(1-B_T^{2}\big)\big(1-\gamma_T\big)}{\big(1-B_R^{2}\big)\big(1-\gamma_R\big)}\Bigg]~.
\end{equation}

Similarly, in the SP the complexity of creating a state with frequency $\omega_{2T}$ starting from a 
state with frequency $\omega_{2R}$ can be written as 
\begin{equation}
\mathcal{C}_{N}^{SP}=\frac{1}{2}\ln \Bigg[\frac{\omega_{2T}}{\omega_{2R}}\Bigg]=\frac{1}{4}\ln \Bigg[
\frac{\big(B_T-1\big)\big(B_T-\gamma_T\big)}{\big(B_R-1\big)\big(B_R-\gamma_R\big)}\Bigg]~.
\end{equation}	
From these expressions, we see that in both cases, the complexity diverges as $\sim\ln|1-B_T|$ if we take the target state to be on 
the transition line keeping the reference state fixed. This is the same as the divergence of the entanglement 
entropy in the LMG model, computed in \cite{Latorre,BDV}. This divergence is clearly due to the presence of the 
``zero mode" at which the harmonic oscillator reduces to the free particle, and is distinct from, say, the 
behaviour of the NC in the transverse XY model where the derivative of the NC (and not the NC itself)
shows a logarithmically diverging behaviour. Here on the other hand, the derivatives of the NC diverge according to
power laws. 

\section{Ground state wavefunction of the LMG model with time-dependent magnetic field}
\label{sec2}

Now we consider the more involved case,
where the external magnetic field is time-dependent, i.e., $B=B(t)$. This variant of the LMG model have been used in 
studying shortcuts to adiabaticity in the context of quantum control \cite{CCPPF}.  
In that case, the LMG model is described by a harmonic oscillator with 
time-dependent frequency. In the first case, when the system is in the BP, the frequency is given by 
$\omega_{1}(t)=2\sqrt{\left(1-B(t)^{2}\right)\left(1-\gamma\right)}$
and in the second case, in the SP, it reads $\omega_{2}(t)=2\sqrt{\left(B(t)-1\right)\left(B(t)-\gamma\right)}$.
The ground state of this model shows a QPT at $B(t)=1$.
To calculate the NC between two ground states of the LMG model with a time-dependent magnetic 
field in the thermodynamic limit, once again we need to know the explicit expressions 
for the position space wavefunctions. For this, we define two sets of positions and momentum operators, 
characterising the system below and above the phase transition, respectively. These are given by the same 
formulas as before (see Eqs. (\ref{coordinatescons})) with $\omega_i$s replaced by $\omega_i(t)$s. 

The quantum theory of a harmonic oscillator with time-dependent frequency is well studied in the literature. 
The solution of the Schrodinger equation for such an oscillator can be obtained by using the   
Lewis-Riesenfeld theory of time-dependent invariant operators, which we generically call $I(t)$ \cite{LR}
(for a text book discussion see \cite{DR}).
In position space, we  can write down the general solution of the time-dependent Schrodinger 
equation for the Hamiltonian in the BP as a weighted linear combination of the eigenfunctions 
of such a time-dependent invariant operator times a phase factor 
\begin{eqnarray}\label{Wavefunction}
\Psi_{1}(Q_{1},t)&=&\sum_{n}c^{n}_{1}e^{i\alpha^{n}_{1}(t)}\phi^{n}_{1}(Q_{1},t)
=\sum_{n}c^{n}_{1}\psi^{n}_{1}(Q_{1},t)~,\nonumber\\
{\rm with}~~~&~&\psi^{n}_{1}(Q_{1},t)=e^{i\alpha^{n}_{1}(t)}\phi^{n}_{1}(Q_{1},t)~,
\end{eqnarray}
where $\phi^{n}_1$ satisfies the eigenvalue equation $I_{1}\phi_1^{n}=\lambda_{1n}\phi_1^{n}$,
with the time independent eigenvalues $\lambda_{1n}$ and $\alpha^{n}_{1}(t)$'s are phase factors,
whose forms will be given below. The functions $\psi^{n}_{1}(Q_{1},t)$  
themselves satisfy the Schrodinger equation for the time-dependent oscillator.

Similarly,  the wave functions corresponding to the Hamiltonian in SP 
(which we call $\Psi_{2}(Q_{2},t)$) can be  written as a linear combination of a second set 
of eigenfunctions $\phi_{2n}$ of a new invariant $I_{2}$. 
These two sets are different because the two Hamiltonians differ from each other in terms of 
time dependent frequencies, and as we shall see, these frequencies determine the auxiliary  functions 
that appear in the invariant operators (hence the wavefunctions). 

The eigenfunctions of the operator $I_{i}(t)$ ($i=1,2$) are given by \cite{DR}
\begin{equation}\label{tdhowave}
\phi^{n}_{i}(Q_{i},t)=\left(\frac{1}{2^{n}n!f_{i}}\right)^{1/2}\exp\left[\frac{i}{2}
\left(\frac{\dot{f}_{i}}{f_{i}}+\frac{i}{f_{i}^{2}}\right)Q_{i}^{2}\right]H_{n}\left(\frac{Q_{i}}{f_{i}}\right)~.
\end{equation}
Here $H_{n}$s are the conventional Hermite polynomials and the time dependent functions 
$f_{i}(t)$ satisfy the following differential equation 
\begin{equation}
\label{auxi}
\frac{d^{2}f_{i}(t)}{dt^{2}}+\omega^{2}_{i}(t)f_{i}(t)-\frac{1}{f_{i}^{3}(t)}=0~.
\end{equation}
To make the invariant operator $I_{i}(t)$ Hermitian, we need to consider only the real solutions 
of Eq. (\ref{auxi}). Clearly the solutions of this differential equation depend on the time dependent 
frequency of the harmonic oscillator which, for our case, is in turn determined by the time dependent 
applied magnetic field. The phase factor $\alpha_{i}(t)$ appearing in  Eq. (\ref{Wavefunction}) 
can be  written in terms of the function $f_{i}$ as 
\begin{equation}
\alpha_{i}^{n}(t)=-\left(n+\frac{1}{2}\right)\int_{0}^{t}\frac{1}{f_{i}^{2}(t^{\prime})}dt^{\prime}~.
\end{equation}
For future reference, we note here that the ground state ($n=0$) wavefunction of the 
LMG model for both the cases considered above is a Gaussian of the form
\begin{equation}\label{n=0}
\psi^{0}_{i}\left(Q_{i},t\right)=\frac{e^{i\alpha^{0}_{i}(t)}}{\sqrt{f_{i}(t)}}\exp\left[-\frac{1}{2} 
\Omega_{i}(t) Q_{i}^{2}\right]~,~~\Omega_{i}(t)=\left(\frac{1}{f_{i}^{2}}-\frac{i\dot{f}_{i}}{f_{i}}\right)~.
\end{equation}
In most physical situation, while solving the above auxiliary equation, we need to fix two conditions on $f(t)$ 
and its derivative on some time instant, say $t_0$. This is usually done by demanding that at $t_0$, the above 
wavefunctions coincides with those of the instantaneous time-independent Hamiltonian $H(t=t_0)$. 
From Eq. (\ref{tdhowave}), we see that the required conditions at $t_0$ are \cite{KA}
\begin{equation}\label{conditionst_0}
f(t_0)=\frac{1}{\sqrt{\omega(t_0)}} \quad \text{and } \quad \dot{f}(t_0)=0~.
\end{equation}

\section{Circuit complexity of the time-dependent LMG model}
\label{sec3}

Now, using the results of \cite{ABHKM} (with the $\mathcal{F}_{2}$ cost functional), the circuit complexity of preparing the 
Gaussian target state written in terms of  the  $Q_{i}$ coordinates
\begin{equation}\label{target}
\Psi^{T}\left(Q_{i},t\right)=\mathcal{N}^{T}_{i}\left(t\right)\exp\Bigg[-\frac{1}{2}\Omega^{T}_{i}(t)Q_{i}^{2}\Bigg]~~,
\end{equation}
starting from the Gaussian reference state written in the same coordinates
\begin{equation}\label{reference}
\Psi^{R}\left(Q_{i},t\right)=\mathcal{N}^{R}_{i}\left(t\right)\exp\Bigg[-\frac{1}{2}\Omega^{R}_{i}(t)Q_{i}^{2}\Bigg]~~,
\end{equation}
is straightforwardly given by 
\begin{eqnarray}\label{complexity}
&~&\mathcal{C}(t)=\frac{1}{2}\sqrt{\left(\ln\frac{\big|\Omega^{T}(t)\big|}{\big|\Omega^{R}(t)\big|}\right)^{2}
+\left(\arctan\left[{\mathcal A}\right]\right)^{2}}~,\nonumber\\
&~&{\mathcal A}=\frac{{\rm Im}[\Omega^{T}(t)]{\rm Re}[\Omega^{R}(t)]-{\rm Re}[\Omega^{T}(t)]{\rm Im}[\Omega^{R}(t)]}
{{\rm Re}[\Omega^{R}(t)]{\rm Re}[\Omega^{T}(t)]+{\rm Im}[\Omega^{R}(t)]{\rm Im}[\Omega^{T}(t)]}~.
\end{eqnarray}
Here, ${\rm Re}$ and ${\rm Im}$ denotes real and imaginary parts respectively, 
$\Omega^{T}(t)$ and $\Omega^{R}(t)$ are respectively the complex frequencies of the target and reference states, 
and $\mathcal{N}^{T}_{i}$ and $\mathcal{N}^{R}_{i}$ are time-dependent normalisation constants. 
The form of these constants will not be important for us.

To calculate the circuit complexity in the two cases considered above, we consider the following situation. 
We apply a time-dependent magnetic field $B(t)$ at $t=0$. When the value of the magnetic field  
reaches the critical value of unity, the phase transition occurs. On the two sides of this critical value,
we have the BP and the SP. In calculating the complexity for 
magnetic field  $0\leq B(t) \leq 1$ (the BP), we take the state $t=0$ to be the reference  
state and the $n=0$ state of case 1 at an arbitrary time $t$  as the target state. For $B(t)>1$, we assume that the 
magnetic field reaches a constant value after some time $t_{1}$ 
where $B(t_{1})>1$  so that we take the state at $t_{1}$ as the target state 
and the $n=0$ state at time $t$ with $B(t)>1$ as the reference.   

\noindent
\textbf{Complexity between two states in the BP:} 
We will, for illustration, take $B(t) = t$ where the dimensions are appropriately adjusted. 
Then, at $t=0$, we have $B(t=0)=0$, so that $\omega_{1}(t=0)=\omega_{10}
=2\sqrt{1-\gamma}$, and we can take the reference 
state in this case to be  Gaussian $\Psi^{R}=\mathcal{N}_{R}\exp\left[-\frac{1}{2}\omega_{10}Q_{1}^{2}\right]$.  
Then, from the expression for the complexity given in Eq. (\ref{complexity}) and the $n=0$ state in Eq. (\ref{n=0}), 
we obtain the exact expression for the circuit complexity $\mathcal{C}_{N}^{SP}(t)=\frac{1}{2}\sqrt{\mP_1^2 + \mQ_1^2}$, with
\begin{equation}\label{comp1} 
\mP_1=\ln\left[\frac{\sqrt{f_{1}^2\dot{f}^{2}_{1}+1}}{\omega_{10}f_{1}^{2}}\right]
~,~~\mQ_1=\arctan\left[f_{1}\dot{f}_{1}\right]~.
\end{equation}
The auxiliary function $f_1(t)$ can be obtained by numerically solving the auxiliary equation (\ref{auxi}) with the conditions  
of Eq. (\ref{conditionst_0}) imposed at the beginning $t_0=0$, i.e. we demand that the wavefunctions of the time-dependent 
oscillator at $t=0$ coincides with those of the instantaneous time-independent  oscillator at $t=0$.  Notice that for the 
magnetic field profile $B(t)=t$  the systems reaches QPT  at $t=1$  and hence up to this time the above formula  
for the complexity is valid. Remarkably, the divergence of the complexity that we saw in the time-independent case
is now replaced by a finite discontinuity. 
That the complexity is everywhere regular can be gleaned from Eq. (\ref{comp1}), where we see that the argument of the logarithm 
does not vanish due to the fact that $f_1$ must be real and non-zero. 

\noindent
\textbf{Complexity between two states in the SP:}  
Now we assume that, moving away from the QPT value $B_c=1$ (reached at $t=1$), the magnetic field 
stops being time-dependent  and acquires a  constant value  $B=2$ at time $t=2$ (these numbers are for illustration only,
and any other choice of constants will be equally valid in our analysis). 
The  state of the system at $t=2$ and at subsequent times is that of a simple harmonic oscillator with frequency   
$\omega_{20}=2\sqrt{2-\gamma}$. Thus we take the target state in this case to be a Gaussian of the form 
$\Psi^{T}=\mathcal{N}_{T}\exp\Big[-\frac{1}{2}\omega_{20}Q_{2}^{2}\Big]$.  
With the reference state to be the $n=0$ state with $B(t)>1$ ($t>1$) we find the complexity 
$\mathcal{C}_{N}^{BP}(t)=\frac{1}{2}\sqrt{\mP_2^2 + \mQ_2^2}$, with
\begin{equation}\label{comp2}
\mP_2=\ln\left[\frac{\omega_{20}f_{2}^{2}}{\sqrt{f_{2}^2\dot{f}^{2}_{2}+1}}
\right]~,~~\mQ_2=\arctan\Big[-f_{2}\dot{f}_{2}\Big]~.
\end{equation} 
In this case to solve for the auxiliary equation we can impose the conditions in Eq. (\ref{conditionst_0}) at $t=2$ where 
the magnetic filed stops being time-dependent and  acquires a constant value such that the wavefunctions at $t=2$ are 
those of the time-independent oscillator at this instant.
From arguments similar to the ones before, we see that the complexity is again regular everywhere, much
like that in the BP.

\begin{figure}
\centering
\includegraphics[width=0.6\linewidth]{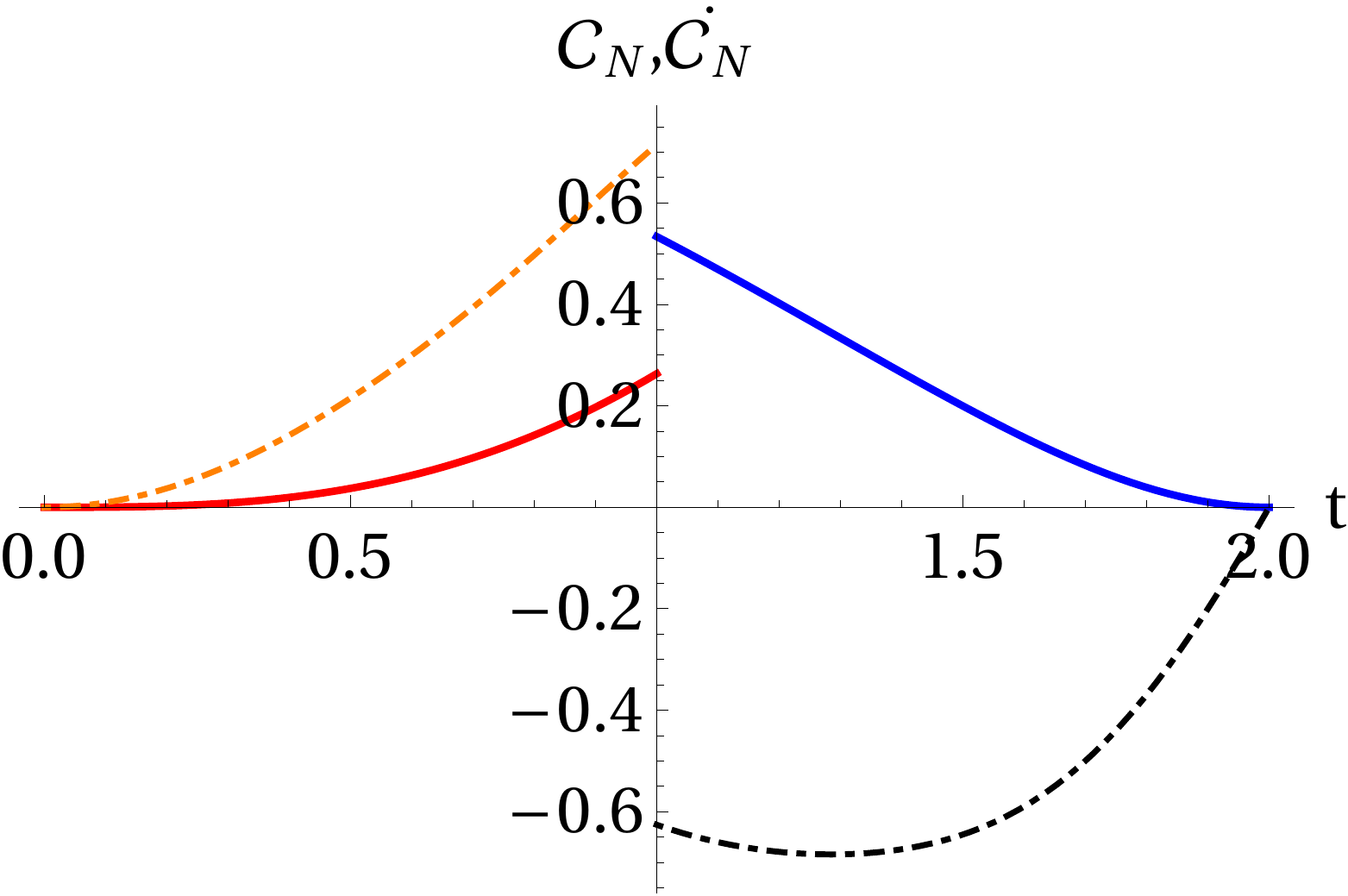}
\caption{Change of Nielsen complexity (solid curves in two phases) and its derivative (dot-dashed curves) across QPT at $t=1$.
We have set $\gamma=0.1$. Both NC and its derivative show discontinuity at QPT. }
\label{fig:complexity-and-derivatives}
\end{figure}
 
In Fig. \ref{fig:complexity-and-derivatives} we have plotted the NC and its derivative with $\gamma=0.1$.
From this plot it can be seen that both NC and its derivative  show discontinuity at QPT. Further at the start ($t=0$)
and the end (at $t=2$) of the time-dependent protocol the NC and its derivative go to zero. 
In the context of comparing the complexity of BP and SP,
it is important to note that the Hamiltonians for cases 1 and 2 have different sets of creation and annihilation 
operators. In the Bogoliubov transformations used to diagonalise these Hamiltonians, the Bogoliubov angles are 
different and do not continuously match to each  other as we change the magnetic field through the transition 
point \cite{CCPPF}. Hence the corresponding position and momentum operators defined in Eq. (\ref{coordinatescons}) are 
also different. This, in particular, means that though for both the cases the the ground state of the system is a Gaussian 
state of the form in Eq. (\ref{n=0}) the coordinates used to write them are different. For this reason, we did 
not calculate the complexity of creating a state from one phase to the other. 

It is however still interesting to ask if the complexity or its derivatives are continuous across the phase transition. 
By this we simply seek to understand whether the NC can be made to approach the same value from both sides, 
when the target states are infinitesimally separated, but in two different phases. The answer is, once  again no. As we 
have shown in Appendix \ref{appfirst}, using the an approximation of the magnetic field close to the QPT and using a general solution 
of the auxiliary equation, the initial conditions imposed upon the auxiliary function force the complexity at 
first order to show discontinuity at QPT. We note that the expressions for the NC 
in two phases that we have written in terms of the auxiliary functions, can also be conveniently expressed in terms of the 
spin correlation functions. This is detailed in Appendix \ref{appexp}.

\section{Fubini-Study complexity of the LMG model}
\label{sec4}

So far we have calculated the circuit complexity of the LMG model ground state using the Nielsen's geometric approach,
where one has to solve the geodesics equations corresponding to the  metric on the space of unitary operators  and 
hence calculate the corresponding minimum distance between two unitary operators defined on this space to find out the circuit complexity. 
An alternative proposal for complexity was put forward in \cite{CHMP}, where one uses the  quantum metric tensor (QMT) 
characterising the distance between two nearby quantum states, rather than the metric on the space of 
unitaries, to calculate the geodesic distance. The resultant complexity is known as the Fubini-Study complexity in the literature.
The mathematical details of the computation of the QMT and the FSC are by now standard in the literature. We will avoid
repeating the details, only the essential definitions are provided in Appendix \ref{QMT}. 

In \cite{SMK}, for the isotropic LMG model, the metric on the space of thermal state was obtained, 
and it was shown that in the limit of zero temperature, the metric components are not well defined. 
The relation between QMT and QPT in the context of LMG model was also studied in \cite{dey}, where 
it was shown that the determinant of the QMT vanishes in the conventional LMG model ground state, 
and hence the inverse is not well defined. 
To overcome these difficulties in the information geometric description of the standard  LMG Hamiltonian of Eq. (\ref{LMGH}), 
recently \cite{GGCHV} proposed a variation of the Hamiltonian of Eq. (\ref{LMGH}), which contains two  
terms linear in the total angular momentum.  Using the HPT, the components of the  QMT of the modified Hamiltonian can
be calculated in the thermodynamic limit $N \rightarrow \infty$. Broadly, it is found that the components of the  metric show 
diverging behaviour at the QPT, however,  the Ricci scalar does not show any particular behaviour at that point.

Concretely, the Hamiltonian we consider here is given by 
\begin{equation}
\label{LMGlinearH}
H =B J_{z}+\Omega J_{x}+\frac{\xi}{j}J_{y}^{2}~,
\end{equation}
and we will mostly be interested in the ground state of this model. Compared to the Hamiltonian of Eq. (\ref{LMGH}), 
here one works with $\gamma=0$ with the addition of an extra term linear in the angular momentum. 
Here, $j=N/2$, and $j(j+1)$ is the eigenvalue of the total spin operator.
 
Since the classical pseudospin operator of the ground state is not along the $z$ axis, before applying the 
Holstein-Primakoff transformations,
we need to perform a rotation about a suitable axis. In this case, the rotation axis is the $y$ axis and the 
rotation angle is $\theta=\arccos\Big[\frac{1}{\sqrt{\Omega^{2}+1}}\Big] $ (see \cite{GGCHV} for details). 
After this rotation - represented by the operator $\mathbf{R}_{y}(\theta)$, is performed, the new Hamiltonian can 
be written in terms of the components of the  
transformed spin operator $J^{\prime}=\mathbf{R}_{y}(\theta) J$ as 
\begin{equation}\label{LMG2tra}
H=\sqrt{\Omega^{2}+1}J^{\prime}_{z}+\frac{\xi}{j}J^{\prime 2}_{y}~.
\end{equation}
Now following a standard procedure, we can obtain the components of the QMT for the Hamiltonian in Eq. (\ref{LMG2tra}) 
by using  the formula given in Eq. (\ref{QGT2}). Among the three parameters in the above Hamiltonian which can be used 
as coordinates in the parameter manifold we take $B=1$ without loss of generality.
The line element of Eq. (\ref{lineelement}) in the $\{\Omega ,\xi\}$ coordinates can be written as 
\begin{equation}\label{QMTLMG}
d\tau^{2}=\frac{\Omega^{2}\left(\mathcal{U}-\mathcal{V}\right)^{2}}{32\mathcal{U}^{2}\mathcal{V}^{4}}
\text{d}\Omega^{2}-\frac{\xi \Omega}{8\mathcal{U}^{2}\mathcal{V}^{2}}\text{d}\Omega\text{d}\xi+\frac{1}{8\mathcal{U}^{2}}\text{d}\xi^{2}~~,
\end{equation}
where we have defined two functions 
\begin{equation}\label{UV}
\mathcal{U}(\Omega, \xi)=\sqrt{\Omega^{2}+1}+2\xi~~~,~~\text{and}~~~ \mathcal{V}(\Omega, \xi)=\sqrt{\Omega^{2}+1}~.
\end{equation}
Importantly, the above line element is slightly different from the one given in \cite{GGCHV}. In particular, 
there is an extra term in the $g_{\Omega \Omega}$ component. The reason for this difference is that we have 
employed the Hamiltonian in Eq. (\ref{LMG2tra}) for computing the metric tensor while on the other hand \cite{GGCHV} 
uses that in Eq. (\ref{LMGlinearH}). The form we have given above is particularly suitable for solving  the geodesic equations 
and hence obtaining the FSC. Note that the metric components  diverge at the QPT, where $\xi=-\frac{1}{2}\sqrt{\Omega^2+1}$,
and this line in the parameter space is called the seperatrix of the model, separating the two coherent states corresponding to ground state.

The metric in Eq. (\ref{QMTLMG}) is non diagonal in the $\{\Omega ,\xi\}$ coordinates so that corresponding geodesics 
equations in terms of them are complicated and are difficult to solve. Our first goal here  is to 
transform to  a set of coordinates in which the metric is diagonal. For this we perform a series of coordinate transformations. 
Since $\xi$ can take both positive and negative values, in the coordinate transformations below we need to consider
two branches $\xi>0$ and $\xi<0$ separately. \\
\noindent
\textbf{Diagonal metric for $\xi>0$ branch:} We first use $\{t=\sqrt{\Omega^{2}+1} ,\xi\}$  
and then transform to $\{t ,y=\xi/t\}$ coordinates, so that  
the line element after these transformations  takes the following  simplified form in the $\{t ,y\}$ coordinates
\begin{equation}\label{line2}
d\tau^{2}=\frac{1}{8t^{2}\left(1+2y\right)^{2}}\Big[y^{2}\text{d}t^{2}+yt\text{d}t\text{d}y+t^{2}\text{d}y^{2}\Big]~.
\end{equation}
The next step is to find another coordinate transformation which will make the above metric diagonal. 
We propose $\{t ,y\}\rightarrow \big\{t=\frac{x_{2}}{\sqrt{x_{1}}}, y=x_{1}\big\}~$.  
The diagonal from of the metric after this transformation, in the $\{x_1,x_2\}$ coordinates is given by 
\begin{equation}\label{line3}
d\tau^{2}=\frac{1}{32x_2^{2}\left(1+2x_1\right)^{2}}\Big[3x_2^{2}\text{d}x_1^{2}+4x_1^{2}\text{d}x_2^{2}\Big]~.
\end{equation}
It is not difficult to check that any transformation of the $t$ coordinate of the form 
$t=\frac{g(x_{2})}{\sqrt{x_{1}}}$, $g(x_{2})$ being an arbitrary function of $x_{2}$ will render the line 
element in Eq. (\ref{line2}) diagonal. We choose to consider the simplest one. 
The Ricci scalar of this metric can be easily calculated and is given by
$R=\frac{128}{3x_{1}}$. For $\xi>0$ we have  $x_1\neq0$, and hence $R$ is regular, which indicates  
that the underlying geometry is well defined everywhere in parameter range that we consider.\\
\noindent
\textbf{Diagonal metric for $\xi<0$ branch:} 
In this case the successive coordinate transformations are the followings. First transform to new set
of coordinates $\{\tilde{t}=\sqrt{\Omega^{2}+1} ,\xi\}$, and  then to the set $\{\tilde{t} ,\tilde{y}=-\xi/\tilde{t}\}$.
Finally the resultant line element in the \{$\tilde{t}$,$\tilde{y}$\} coordinates can be made diagonal by transforming to
$\big\{\tilde{t}=\frac{\tilde{x}_{2}}{\sqrt{\tilde{x}_{1}}}, \tilde{y}=\tilde{x}_{1}\big\}~$.  The metric after these 
transformations is given by
\begin{equation}
\label{line4}
d\tau^{2}=\frac{1}{32 \tilde{x}_2^{2}\left(1-2 \tilde{x}_1\right)^{2}}\Big[3\tilde{x}_2^{2}\text{d}\tilde{x}_1^{2}
+4\tilde{x}_1^{2}\text{d}\tilde{x}_2^{2}\Big]~,
\end{equation}
By comparing with the previous  case we see that here only the second transformation is different. 
Since $\xi<0$, $\tilde{y}$ and hence $\tilde{x}_1$ are positive. Notice also that when $\xi<0$ the coordinates 
$x_1$ and $x_2$ defined previously would become invalid.
The Ricci scalar computed from this metric is given by $R=-\frac{128}{3\tilde{x}_1}$ is regular at QPT where 
the separatrix equation reads $\tilde{x}_1=1/2$.
Importantly, we note from Eq. (\ref{inversetra2}) that we have to restrict ourselves to the region
$\tilde{x}_2^2>\tilde{x}_1$ to avoid the coordinate $\Omega$ from becoming imaginary.
We will be careful about this in what follows. 

\subsection{Computing the FSC in the LMG model}

We now calculate the FSC for metrics obtained above by minimising the complexity functional given in Eq. (\ref{FScost}). 
The task is to solve the geodesic equations for the QMTs obtained above, and then to find out the minimum distance 
between two points given by the coordinates $\{\Omega_1,\xi_1\}$ and $\{\Omega_2,\xi_2\}$ on the parameter manifold. 
The mathematical details are somewhat cumbersome, and are given in Appendix \ref{AppFS}. \\
\noindent
\textbf{FSC for $\xi>0$.} 
The geodesic equations corresponding to QMT of Eq. (\ref{line3}) can be separated using the Hamilton-Jacobi method,
the details of which are given in Appendix  \ref{AppFS}, from which we quote the results of Eq. (\ref{ComFSMain}),
\begin{equation}
\begin{split}
\mathcal{C}_{FS}
=\sqrt{\frac{3}{2}}\frac{1}{8}\ln\Bigg[\frac{1}{\xi_{1}}\Bigg\{\frac{2\xi_2^{2}+
\xi_1\sqrt{\Omega_{1}^{2}+1}}{2\xi_1+\sqrt{\Omega_{1}^{2}+1}}\Bigg\}\Bigg].
\end{split}
\end{equation}
This is the expression for the complexity in terms of the boundary point coordinates. 
Importantly, as explained in Appendix  \ref{AppFS},
the complexity does not depend on $\Omega_{2}$, rather $\Omega_{2}$ is itself determined by other constants.\\
\noindent
\textbf{FSC for $\xi<0$.} 
An entirely similar analysis holds for the more interesting case $\xi<0$, and we find here 
\begin{equation}
\label{ComFS-xi<0}
\begin{split}
\mathcal{C}_{FS}=\sqrt{\frac{3}{2}}\frac{1}{8}\ln\Bigg[\frac{\xi_{1}\big(2\xi_1+\sqrt{\Omega_{1}^{2}+1}\Big)
}{\xi_1\sqrt{\Omega_{1}^{2}+1}+2\xi_2^{2}}\Bigg].
\end{split}
\end{equation}
With this expression we can study the behaviour of FSC close to the QPT.
Suppose that we move the initial  point $\{\Omega_1,\xi_1\}$ of the geodesic close to the separatrix
$\xi=-\frac{1}{2}\sqrt{\Omega^{2}+1}$. 
In this limit the complexity diverges, indicating the QPT. Thus unlike the Ricci scalar of the QMT, the complexity detects the QPT.
It is also not difficult to show that the FSC of the excited states of the LGM model can be computed in 
the same way as above, and we have provided the details in Appendix \ref{LMG2app}. As in this case, the FSC of the
excited states also shows non-analyticity at the phase transition. 

\begin{figure}[h!]
\centering
\includegraphics[width=0.6\linewidth]{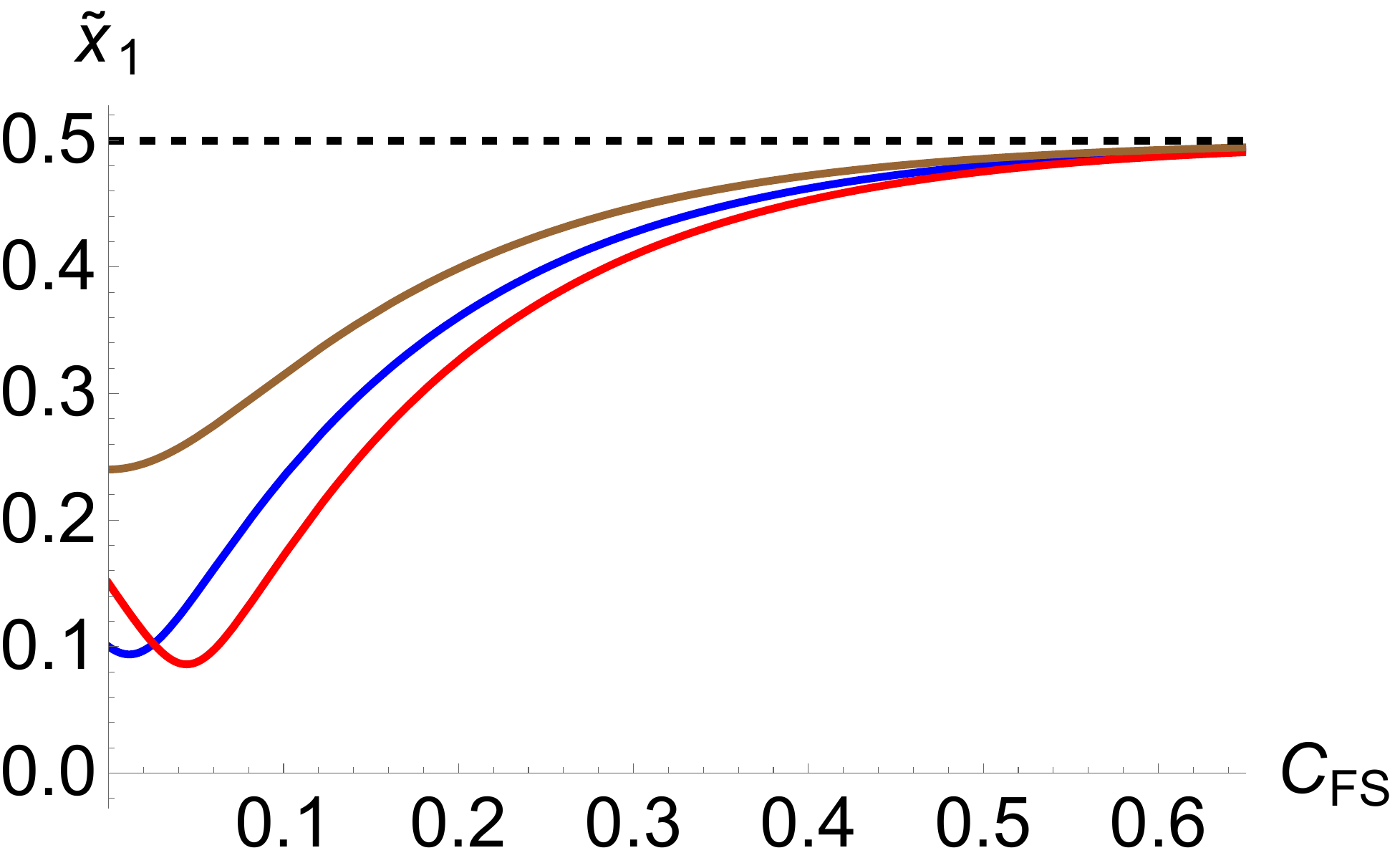}
\caption{Numerical solution of $\tilde{x}_1$ as a function of $\tau$. The blue, red and brown curves correspond to the 
initial conditions $(\tilde{x}_1,\tilde{x}_2,\dot{\tilde{x}}_1)=(0.1,0.32,-1)$, $(0.15,0.42,-2)$ and $(0.24,0.5,9)$ respectively. The horizontal
black dashed line denotes $\tilde{x}_1=\frac{1}{2}$.}
\label{lmgn1}
\end{figure}
It is important to quantify the divergence of the FSC near the separatrix. For this, it is more convenient to 
describe geodesics that we shoot from a given reference point away from the separatrix. Here, we solve the geodesic 
equations resulting from Eq. (\ref{line4}) numerically.
Numerical solutions of $\tilde{x}_1$ and $\tilde{x}_2$ are obtained as functions of the affine parameter, given interpolating functions which are inverted using a standard root finding procedure in Mathematica
to find the FSC. Fig. \ref{lmgn1} shows the numerical solutions of $\tilde{x}_1$ as a function of the affine parameter (which is the FSC itself). In this figure, the blue, red and brown curves correspond to the 
initial conditions $(\tilde{x}_1,\tilde{x}_2,\dot{\tilde{x}}_1)=(0.1,0.32,-1)$, $(0.15,0.42,-2)$ and $(0.24,0.5,9)$ respectively.
It is clearly seen that, even with an initial velocity vector that moves a geodesic away from the  separatrix 
(the horizontal black dashed line), it is eventually  attracted towards it  and reach there
``slowly," i.e., near this line, a finite change in the affine parameter produces an infinitesimal change in $x_1$. This is analogous
to what happens in dynamical systems when trajectories are close to the separatrix. Also, we
have checked that $\tilde{x}_2^2>\tilde{x}_1$ all along the geodesic, as required. 

\begin{figure}[h!]
\centering
\includegraphics[width=0.6\linewidth]{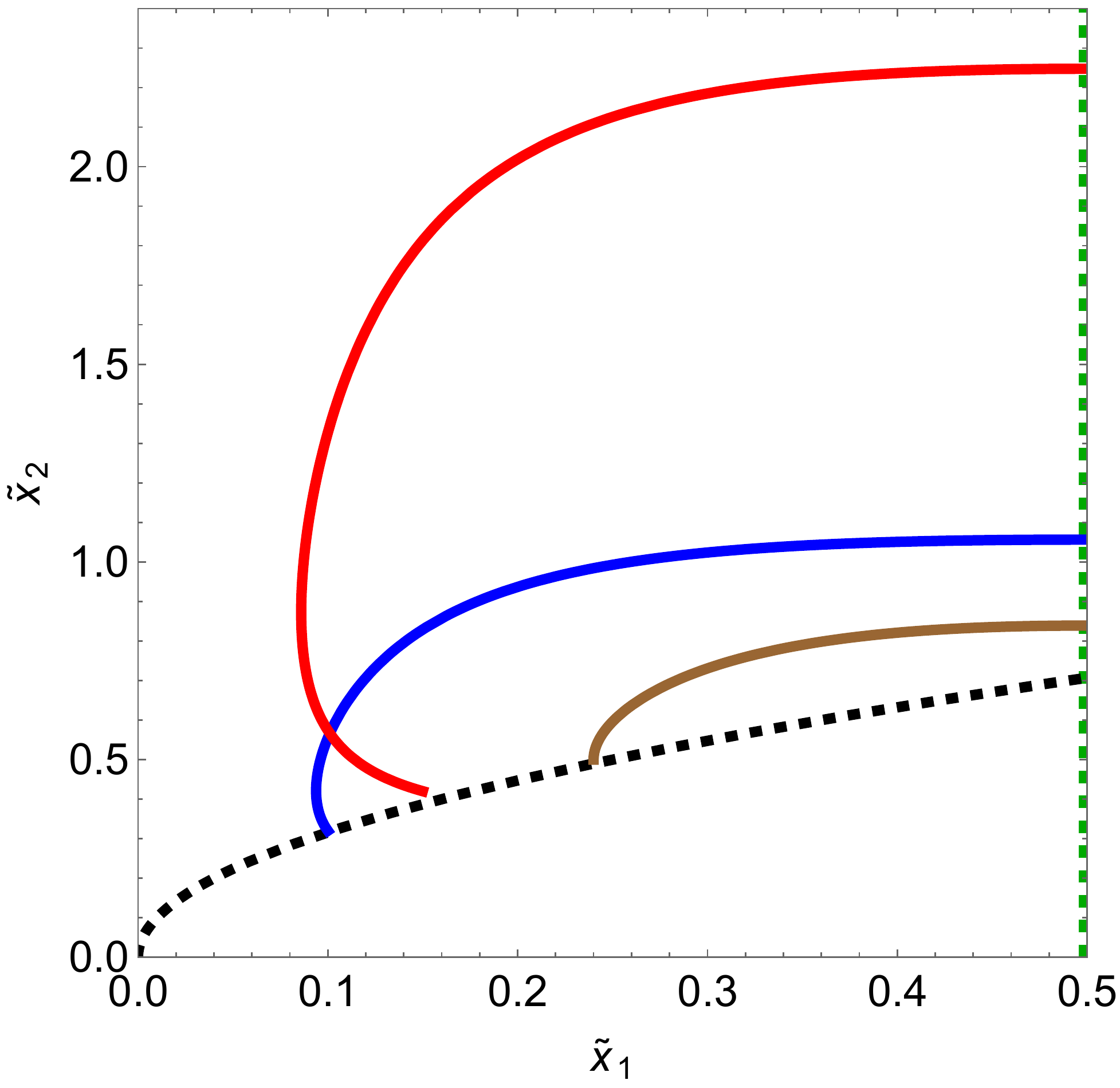}
\caption{Geodesics on the $\tilde{x}_1-\tilde{x}_2$ plane, the dashed black line here represents
	 $\tilde{x}_2^2-\tilde{x}_1=0$, and the vertical 
dashed green line is the separatrix $\tilde{x}_1=\frac{1}{2}$.}
\label{lmgn2}
\end{figure}

\begin{figure}[h!]
\centering
\includegraphics[width=0.6\linewidth]{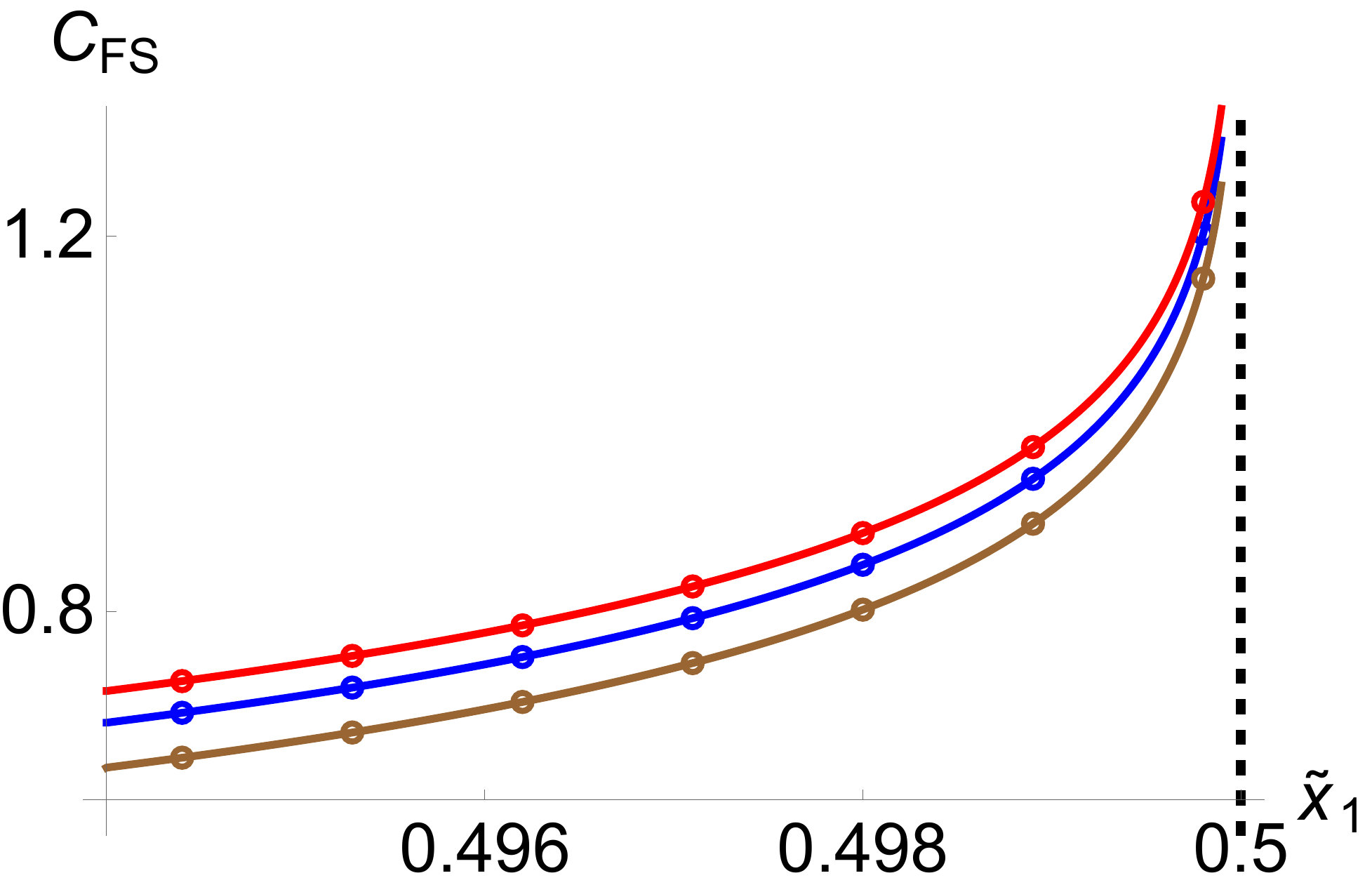}
\caption{The FSC of Fig. \ref{lmgn1} fitted with the function $a + b\ln(0.5-\tilde{x}_1)$. The  colour coding is the same as before.
For the blue, red and brown curves, $a=-0.1, -0.07$ and $-0.15$ respectively, and $b=-0.153$ in all cases.
The vertical dashed black line is the separatrix $\tilde{x}_1=\frac{1}{2}$.}
\label{lmgn3}
\end{figure}

Fig. \ref{lmgn2} shows these geodesics in the $\tilde{x}_1-\tilde{x}_2$ plane, where the same colour coding as in 
Fig. \ref{lmgn1} is followed. We have drawn the geodesics as parametric plots on a flat $\tilde{x}_1-\tilde{x}_2$ plane, but these
carry the same information as one would obtain if these are drawn on the curved parameter manifold embedded in three dimensions. 
Clearly, all geodesics bend towards the separatrix, confirming our assertion in
the last paragraph. The dashed black line here represents $\tilde{x}_2^2-\tilde{x}_1=0$,
and the vertical dashed green line is the separatrix. 
Finally, in Fig. \ref{lmgn3}, we show numerical fits of the FSC near the separatrix with the 
function $a + b\ln(0.5-\tilde{x}_1)$, where the same colour 
coding of Fig. \ref{lmgn1} has been used. The blue, red and brown curves are fitted with 
$a=-0.1, -0.07$ and $-0.15$ respectively. The solid lines correspond to the numerical solutions and the circles of the
same colours denote the fits. In all cases, we find $b=-0.153$, in precise agreement with the
pre-factor of Eq. (\ref{ComFS-xi<0}). 
The computation of the FSC for excited states in the LMG model of Eq. (\ref{LMGlinearH}) is also straightforward.
The details of this computation are presented in Appendix \ref{LMG2app}. 

\subsection{Nielsen complexity of the modified LMG model ground state}

Before concluding this section we shall briefly discuss the calculation of the Nielsen complexity of this modified 
LMG model.  Starting from the rotated form of the Hamiltonian 
of the modified LMG model (in Eq. (\ref{LMG2tra})),  first applying the Holstein-Primakoff transformations and then 
performing a subsequent Bogoliubov transformation  to diagonalise  it the final form of the Hamiltonian in the 
ground state can be written as \cite{GGCHV}
\begin{equation}
\mathbf{H}_{LMG2}\approxeq-j\mathcal{V}(\omega,\xi)+\sqrt{\mathcal{U}(\omega,\xi)\mathcal{V}(\omega,\xi)}
\Big(b_{0}^\dagger b_{0}+\frac{1}{2}\Big)~,
\end{equation}
where $b_{0}$ and $b_{0}^\dagger$ are corresponding creation and annihilation operators after the BT and 
$\mathcal{U},\mathcal{V}$ are defined before in Eq. (\ref{UV}). Once again this can recognised as  the Hamiltonian 
of  a simple harmonic oscillator with frequency $\omega_{m}=\sqrt{\mathcal{U}(\omega,\xi)\mathcal{V}(\omega,\xi)}$. 
It is straightforwardly shown that the circuit complexity of creating a ground state with frequency $\omega_{m2}$  
starting from a ground state $\omega_{m1}$ is given by
\begin{equation}
\mathcal{C}_{N}=\frac{1}{2}\ln\bigg[\frac{\omega_{m2}}{\omega_{m1}}\bigg]=\frac{1}{4}\ln
\Bigg[\frac{\Omega_{2}^{2}+1+2\xi_{2} \sqrt{\Omega^{2}_{2}+1}}{\Omega^{2}_{1}+1+2\xi_{1} \sqrt{\Omega^{2}_{1}+1}}\Bigg]~.
\label{NCLMGnew}
\end{equation}
The NC shows the same divergence structure as the FSC, as is easily gleaned from Eqs. (\ref{ComFS-xi<0}) and (\ref{NCLMGnew}). 

\section{Conclusions and discussions}
\label{concl}

Complexity has now become ubiquitous in the study of quantum field theories, and one can gain deep insights 
by investigating its behaviour in simpler spin systems. So far in the literature, the computation of complexity 
has mostly focussed on the transverse XY model and its variants. These models 
have a non-trivial phase diagram, but are simple in the sense that they are reducible to quasi free fermionic systems 
and the NC as well as the FSC are related to the Bogoliubov angle. Here we have investigated the LMG model which contains 
long range interactions, and hence does not share this property. We found remarkably different features of the
complexity here, compared to the XY model. 

While for a time-independent scenario, the NC and the FSC diverges at the phase transition, the time-dependent
NC avoids this singularity. The divergence in the former cases can be traced back to the presence of a 
zero mode, quantified by the separatrix of the model. Importantly, this divergence is the same as the one 
in a computation of the entanglement entropy of the model. In the latter case, we applied the Lewis-Riesenfeld method of 
obtaining the wavefunctions of such non stationary systems to show why the NC of the LMG model avoids this singularity. 
These results are novel and to the best of our knowledge have not appeared in the literature before.  

As a future direction, it will be very interesting to compute the FSC in a quench scenario, as was done
recently for the quenched XY model \cite{JGS2}. The non-trivial nature  of the information
metric probably holds the clue to the behaviour of the FSC during the quench. We leave such a study for the future. 
Furthermore it will be interesting to extend such studies in other models such as the Dicke model of quantum optics.

\appendix
\section{Complexity of the LMG model close to QPT}
\label{appfirst}

For a given time-dependent magnetic field profile in sec-\ref{sec2} we numerically solved the auxiliary equation to find 
out the complexity and observed that it shows discontinuous behaviour at QPT.  In this appendix we shall discuss the 
behaviour of the Nielsen  complexity when the reference and the target state are very close to the QPT,  using the 
general solution of the auxiliary equation written in terms of the solution of the classical equation of motion 
of the time-dependent oscillator. 
We shall argue that, in general, the complexity is not contentious across the QPT, i.e. its numerical value in the 
BP and the SP are different at QPT, in agreement with the discussions of the  main text.

Since at the QPT  the magnetic field approaches the critical value $B_{c}=1$, we can introduce a new variable $\epsilon$  
and at the first order approximation can write the magnetic field before and after the phase transition 
(i.e., in the BP and SP respectively) as 
\begin{equation}
\label{Bapprox}
B_{1}(\epsilon)\approx 1-\epsilon~,~~\text{and}~~B_{2}(\epsilon) \approx 1+\epsilon~,
\end{equation}
where the QPT is understood to be reached in the limit $\epsilon \rightarrow 0$. Here, $\epsilon$ is assumed to be varied over a small 
range around zero i.e., before and after QPT. In this setting the variable $\epsilon$ plays the role of $t$ in section \ref{sec2}.
	
With this approximation, the $\epsilon$ dependent frequencies in the two phases  can be written up to first order as
\begin{eqnarray}
\label{appfrequency}
\omega_{1}(\epsilon)\approx 2\sqrt{2\epsilon (1-\gamma)}=b_{1}\sqrt{\epsilon}~,~~\nonumber\\
\omega_{2}(\epsilon)\approx 2\sqrt{\epsilon (1-\gamma)}=b_{2}\sqrt{\epsilon}~,
\end{eqnarray}
where we have conveniently introduced the constants $b_1$ and $b_2$. 
Thus very close to the QPT, we have an $\epsilon$ dependent oscillator (instead of a time dependent oscillator 
of section \ref{sec2}), with the frequencies at the both phases varying as $\sqrt{\epsilon}$, 
and their respective magnitudes being different. 
This makes the analysis much simpler - we need to perform the calculations only in one of the two phases and the 
result for the other phase can be obtained  by changing the constant $b_{i}$.
	
To obtain the complexity, we need to find the solution of the auxiliary equation Eq. (\ref{auxi}), with the 
variable $\epsilon$. This can be obtained using the following procedure. Suppose $g_{1}(\epsilon)$ and $g_{2}(\epsilon)$ 
are two linearly independent solutions of the classical equation of motion of the variable frequency oscillator
\begin{equation}\label{classical}
	\frac{d^{2}g(\epsilon)}{d\epsilon^{2}}+\omega^{2}(\epsilon)g(\epsilon)=0~.
\end{equation}
If we denote by $W$ their Wronskian then a solution of the corresponding auxiliary 
equation can be written as 
\begin{equation}
f(\epsilon)=\sqrt{A_1g_1^2(\epsilon)+2A_2g_1(\epsilon)g_2(\epsilon)+2A_3g_2^2(\epsilon)}~,
\end{equation}
with the three consonants $A_1,A_2, A_3$ being related by the relation
\begin{equation}
A_1A_3-A_2^2=\pm\frac{1}{W^2}~.
\end{equation}
The two independent   constants $A_1$ and $A_3$ can be obtained by using the conditions listed in 
Eq. (\ref{conditionst_0}) at a value $\epsilon_1$ in the BP and $\epsilon_2$ in the SP, with both of them 
being very much smaller than unity i.e. two points being very close to QPT in the respective phases. Notice 
also that the constant $A_i$ take different values in two phases (since the conditions of the 
auxiliary functions are different).

Two linearly independent solutions of the classical equation of motion Eq. (\ref{classical}) with the frequency  
in Eq. (\ref{appfrequency}) are the Bessel functions given respectively by
\begin{equation}
g_{1}(\epsilon)=\epsilon^{1/2}J_{1/3}(y)~,~g_{2}(\epsilon)=\epsilon^{1/2}J_{-1/3}(y)~;~y=\frac{2b_i}{3}\epsilon^{3/2}.
\end{equation}
Here $b_i$ stands for either $b_{1}$ or $b_{2}$.

With this solution, we can now calculate the limiting values at the QPT ($\epsilon  \rightarrow 0 $) of 
the quantities of our interest. Firstly, using the standard expansion of the Bessel functions for small values 
of the argument, we obtain the value of the auxiliary function at the QPT to be 
$f(\epsilon \rightarrow 0)=\frac{\sqrt{A_{i1}}}{\Gamma(2/3)}\big(\frac{3}{b_i}\big)^{1/3}$. Since $b_{1}\neq b_{2}$, 
and $A_1$ in two phases are different
the auxiliary function in general  is not continuous at the phase transition. Similarly in the limit 
$\epsilon \rightarrow 0 $, the derivative of the auxiliary function is 
$\frac{df(\epsilon)}{d\epsilon} \rightarrow nb_i^{1/3}\sqrt{\frac{27 A_{i1} A_{i3}-4\pi^2}{A_{i1}}}$, 
$n$ being a constant. 

Now we can substitute these values in the expression for the complexity to obtain the following  
first order formula in either of the two phases
\begin{equation}
\begin{split}
&\mathcal{C}_i(\epsilon \rightarrow 0)=\frac{1}{2}\sqrt{\mathcal{P}_{i0}^{2}+\mathcal{Q}_{i0}^{2}}~; \\
&\mathcal{P}_{i0}=\ln \bigg[\bigg(\frac{3}{b}\bigg)^{2/3}\frac{A_{i1}\omega_{i0}}{\sqrt{1+3^{2/3}n^2(27 A_{i1} A_{i3}-4\pi^2)}}\bigg]~,\\
&\mathcal{Q}_{i0}=\arctan \bigg[\frac{3^{1/3}}{\Gamma(2/3)}n\sqrt{27 A_{i1} A_{i3}-4\pi^2}\bigg]~.
\end{split}
\end{equation}
This expression shows that  complexities in both the phases, with first order values of $f(\epsilon)$ 
and $df/d\epsilon$, are different  from one another in the limit $\epsilon \rightarrow 0$, and hence discontinuous at QPT. 

Here it is important to recognize the important role played by the conditions on the auxiliary function $f(\epsilon)$ 
at $\epsilon_1$ (and $\epsilon_2$). If these conditions were not imposed upon $f(\epsilon)$ then it is possible to 
construct a solution such that the complexity  and its derivative is continuous across the QPT. However, if we impose 
restriction on the auxiliary function at some initial  (or final) point of evolution then the solution necessarily 
changes in such a way that value of the complexity becomes discontinuous at QPT. Furthermore, these conclusions are 
valid only when the reference and the target states are chosen to be close to the QPT. In the general case the 
conclusions of sec-\ref{sec2}  remain valid.	

\section{Circuit Complexity of the LMG model in terms of the uncertainties}
\label{appexp}

In the basis of the eigenfunctions of the invariant operator we have the
following diagonal matrix elements
\begin{equation}
\begin{split}
&\big<\phi_{n}(t)\big|Q_{i}\big|\phi_{n}\big>=\big<\phi_{n}(t)
\big|P_{i}\big|\phi_{n}\big>=0~,\\
&\big<\phi_{n}(t)\big|Q^{2}_{i}
\big|\phi_{n}\big>=f_{i}^{2}\Big(n+\frac{1}{2}\Big)~,\\
&\big<\phi_{n}(t)\big|P^{2}_{i}\big|\phi_{n}\big>=\frac{1}{f_{i}^{2}}
\Big(1+\dot{f}_{i}^{2}f_{i}^{2}\Big)\Big(n+\frac{1}{2}\Big)~,\\      
&\big<\phi_{n}(t)\big|P_{i}Q_{i}+Q_{i}P_{i}
\big|\phi_{n}\big>=2f_{i}\dot{f}_{i}\Big(n+\frac{1}{2}\Big)~.
\end{split}
\end{equation}
Hence the circuit complexity  calculated above in either of the two phases
can be written as ${\mathcal C}_i(t)=\frac{1}{2}\sqrt{{\mathcal A_i^2+{\mathcal B}_i^2}}$, with
\begin{equation}
\label{expectations}
\begin{split}
\mathcal{A}_{i}(t)=\ln\Bigg[\frac{\big<\Delta
P_{i}\big>_{T}}{\big<\Delta
P_{i}\big>_{R}}\frac{\big<\Delta
Q_{i}\big>_{R}}{\big<\Delta
Q_{i}\big>_{T}}\Bigg]~,
\mathcal{B}_{i}(t)=\arctan\Big[\frac{1}{2}\big<P_{(i}Q_{i)}\big>_T\Big]~,
\end{split}
\end{equation}
where $\big<\Delta Q_{i}\big>_{T}$ and  $\big<\Delta
P_{i}\big>_{T}$   respectively indicates the uncertainty of the
position and the  momentum operator in the target state in either phase.
Similarly a subscript `R' indicates corresponding quantities in the
reference state.  After a lengthy algebraic calculation this expression
can be rewritten in terms of the original total spin operators of that
appear LMG model Hamiltonian. We first present the result in the symmetric phase, 
for which the expression is relatively simpler, and is given by
\begin{equation}
\label{complexity-uncertatinity}
\begin{split}
\mathcal{A}_{2}(t)=\frac{1}{2}\ln\Bigg[\frac{e^{2\Theta_{2}}
\omega_{2}^2(t)}{\omega_{20}^2}\frac{\big<J_{y}^{2}\big>_{T}}{\big<J_{x}^{2}
\big>_{T}}\Bigg]~,
{\mathcal B}_2(t)=\arctan\Big[\frac{1}{N}\big<J_{(x}J_{y)}
\big>\Big]~,
\end{split}
\end{equation}
where $\Theta_{2}$ is the angular  variable  in the Bogoliubov
transformations and is given by $\Theta_{2}=\text{arctanh}
\big((1-\gamma)/(2B(t)-1-\gamma)\big)$. Also,
$J_{(x}J_{y)}=\frac{1}{2}(J_{x}J_{y}+J_{y}J_{x})$.
On the other hand, in the broken
phase the same expression is true but with the rotated spin components
$\tilde{S}_{i}$s instead of the original ones. After the
appropriate transformations are made, the expression for the complexity reads
\begin{equation}
\begin{split}
&\mathcal{A}_{1}(t)=\frac{1}{2}\ln\Bigg[\frac{e^{2\Theta_{1}}
\omega_{1}^{2}(t)\big<J_{y}^{2}\big>_{T}}{\omega_{10}^{2}Y}\Bigg]~,\\
&\mathcal{B}_{1}(t)=\arctan\Big[\frac{\cos^{2}\theta_{0}}{2N}\big<J_{(x}J_{y)}\big>\Big]~,
\end{split}
\end{equation}
where
\begin{equation}
Y=\cos^{2}\theta_{0}\big<J_{x}^{2}\big>_{T}+\sin^{2}\theta_{0}
\big<J_{z}^{2}\big>_{T}-\sin\theta_{0}\cos\theta_{0}\big<J_{(z}J_{x)}\big>~,
\end{equation}
$\theta_{0}$ is the rotation angle, and $\Theta_{1}=\text{arctanh}\big((B^{2}(t)-\gamma)/(2-B^{2}(t)-\gamma)\big)$ 
is the angular variable used in the Bogoliubov transformation used in this phase.

When the magnetic field becomes time-independent, then $f(t)=\text{constant}$, and
hence from Eq. (\ref{expectations}) we see that the second term in the
expression (\ref{complexity-uncertatinity}) vanishes and we get back the
results of section \ref{sec1} written in terms of the matrix elements of $P^2$ or
$Q^2$.

\section{The quantum metric tensor}
\label{QMT}

To compute the QMT, we consider the Hamiltonian operator  to be parameterised by $k$ real parameters 
$\{x^{i}\}$, $(i=1,2,\cdots,k)$. These parameters naturally define a manifold $\mathcal{M}_{k}$ with coordinates 
$x=(x^{1}, x^{2},\cdots,x^{k})$. Then, following \cite{PV}, we define the rank two covariant tensor known 
as the quantum geometric tensor (QGT)
\begin{equation}\label{QGT}
\mathcal{Q}_{ij}=\bigg<\frac{\partial n(x)}{\partial x^{i}}\bigg|\frac{\partial n(x)}{\partial x^{j}}
\bigg>-\bigg<\frac{\partial n(x)}{\partial x^{i}}\bigg|n(x)\bigg>\bigg<n(x)
\bigg|\frac{\partial n(x)}{\partial x^{j}}\bigg>~,
\end{equation}
where $\big|n(x)\big>$ denotes the $n$'th eigenstate of the Hamiltonian with energy $E_{n}(x)$ i.e., 
$H(x)\big|n(x)\big>=E_{n}(x)\big|n(x)\big>$. In defining the QGT, we have made the reasonable assumption 
that the Hamiltonian and hence the eigenstates as well as the energies are smooth functions of all the parameters.

In general, the components of the QGT are complex functions of the parameters. The real part, a symmetric  
covariant tensor of rank two naturally defines a Riemannian structure on the space of states and 
we call this tensor the QMT (the imaginary part is the Berry curvature)
\begin{equation}
g_{ij}=\Re\big[\mathcal{Q}_{ij}\big]~,
\end{equation}
so that the distance between two quantum states $|n(x)\big>$ and $|n(x+dx)\big>$ differing infinitesimally   
in the parameters can be written as
\begin{equation}\label{lineelement}
d\tau^{2}=1-\big|\big<n(x)\big|n(x+dx)\big>\big|^{2}=g_{ij}\text{d}x^{i}\text{d}x^{j}+\mathcal{O}\left(|dx|^{3}\right)~.
\end{equation}
By using the standard formula for the overlap between a state and its derivative  
$\big<n(x)\big|\frac{\partial n(x)}{\partial x^{i}}\big>$ in terms of the expectation value of 
derivatives of the Hamiltonian itself one can obtain an alternative formula for the QGT given as
\begin{equation}\label{QGT2}
\mathcal{Q}_{ij}=\sum_{m\neq n}\frac{\big<n(x)\big|\partial_{i}H\big|m(x)\big>\big<m(x)
\big|\partial_{j}H\big|n(x)\big>}{\left(E_{m}(x)-E_{n}(x)\right)^{2}}~,
\end{equation}
with the subscript $i$ denoting the partial derivatives with respect to $x^{i}$ (see \cite{Gu} 
for a derivation of this formula as well as for a review of approach of using QGT to detect QPTs).
We shall use this form of the QGT to calculate the QMT of the LMG model.

Let us consider a path $x^{a}(\tau)$ on the space characterize by QMT.
Then, with the QMT calculated from the above formula in Eq. (\ref{QGT2}),
the FSC is obtained by minimising the quantity
\begin{equation}\label{FScost}
\mathcal{C}_{FS}=\min\int_{0}^{1}d\tau\sqrt{g_{ab}\frac{dx^{a}}{d\tau}\frac{dx^{b}}{d\tau}}~~.
\end{equation}
Thus the complexity between two states on the parameter manifold is given by the minimum value of the 
distance of the geodesic connecting these two points.
In writing the above formula we have assumed that the geodesic $x^a(\tau)$  connecting  reference state and target state 
on the parameter manifold is 
parameterized in such a way that reference state corresponds to $\tau=0$, while the target state is at $\tau=1$.

\section{Calculation of the Fubini-Study complexity in the LMG model}
\label{AppFS}

\noindent
\textbf{Separation of geodesic equations for  $\xi>0$:} 
First we write down the relations between $\{\Omega,\xi\}$ and $\{x_1,x_2\}$ coordinates 
\begin{equation}
x_1=\frac{\xi}{\sqrt{\Omega^{2}+1}}~, ~ x_2=(\Omega^{2}+1)^{1/4}\sqrt{\xi}~,
\end{equation}
with the inverse relations being given by 
\begin{equation}\label{inversetra}
\xi=\sqrt{x_1}x_2~, ~\Omega=\sqrt{\frac{x_2^2}{x_1}-1}~.
\end{equation}

The geodesic equations corresponding to the metric in Eq. (\ref{line3}) can be easily written down and these are given by 
\begin{eqnarray}
\frac{3}{16}\frac{d}{d \tau}\Bigg(\frac{\dot{x}_{1}}{\big(1+2x_1\big)^{2}}\Bigg)
-\frac{\Big(2x_{1}\dot{x}_{2}^{2}-3\dot{x}_{1}^{2}x_{2}^{2}\Big)}{8x_{2}^{2}\big(1+2x_1\big)^{3}}=0~,\nonumber\\
\frac{d}{d \tau}\Bigg(\frac{x_{1}^{2}\dot{x}_{2}}{x_{2}^{2}\big(1+2x_1\big)^{2}}\Bigg)
+\frac{x_{1}^{2}\dot{x}_{2}^{2}}{x_{2}^{3}\big(1+2x_1\big)^{2}}=0~.
\label{geoLMG}
\end{eqnarray}
These equations are difficult to solve analytically. Instead of directly solving these second order nonlinear differential equations,  
we shall use the well known Hamilton-Jacobi (HJ) method for separating them, so that we obtain first order equations. 
As we shall see, there exists a hidden integral of motion which makes the separation possible. 
The HJ equation,  
\begin{equation}
2\frac{\partial S(x^{a},\tau)}{\partial \tau}=g^{ab}\Bigg(\frac{\partial S(x^{a},\tau)}{\partial x^{a}}\Bigg)
\Bigg(\frac{\partial S(x^{a},\tau)}{\partial x^{b}}\Bigg)~~,
\end{equation}
with the principal function denoted as $S(x^{a},\tau)$, for the metric in Eq. (\ref{line3}) is given by
\begin{equation}
\begin{split}
2\frac{\partial S(x^{a},\tau)}{\partial \tau}&=\frac{32}{3}\left(1+2x_1\right)^{2}
\Bigg(\frac{\partial S(x^{a},\tau)}{\partial x_{1}}\Bigg)^{2}\\
&+8\frac{x_{2}^{2}}{x_{1}^{2}}
\left(1+2x_1\right)^{2}\Bigg(\frac{\partial S(x^{a},\tau)}{\partial x_{2}}\Bigg)^{2}~.
\end{split}
\end{equation}
Now as is the standard procedure of separation, we assume the following form of the principal function
\begin{equation}
 S(x_{1},x_{2},\tau)=\frac{1}{2}K^{2}\tau+S_{x_{1}}\left(x_{1}\right)+S_{x_{2}}\left(x_{2}\right)~,
\end{equation}
where, as indicated, $S_{x_{1}}\left(x_{1}\right)$ and $S_{x_{2}}\left(x_{2}\right)$ are two functions 
of their single arguments, and $K^{2}=g_{ab}\dot{x}^{a}\dot{x}^{b}$ is a constant along 
the geodesic. Substituting this form of the principle function in the HJ equation, we get
\begin{equation}
K^{2}=\frac{32}{3}\left(1+2x_1\right)^{2}\Bigg(\frac{d S_{x_{1}}}{d x_{1}}\Bigg)^{2}+8\frac{x_{2}^{2}}{x_{1}^{2}}
\left(1+2x_1\right)^{2}\Bigg(\frac{dS_{x_{2}}}{d x_{2}}\Bigg)^{2}~~.
\end{equation}
After a bit of algebraic manipulation this can be rewritten as 
\begin{equation}\label{separate}
\frac{x_{1}^{2}}{\left(1+2x_1\right)^{2}}K^{2}-\frac{32}{3}x_{1}^{2}\Bigg(\frac{d S_{x_{1}}}{d x_{1}}\Bigg)^{2}
=8x_{2}^{2}\Bigg(\frac{dS_{x_{2}}}{d x_{2}}\Bigg)^{2}=\mathcal{M}~,
\end{equation}
with ${\mathcal M}$ being a constant. 
The left side of Eq. (\ref{separate}) is a function of $x_{1}$ only, and the right side depends only 
on $x_{2}$, and hence for the above equality to be  valid, both sides must be a constant as indicated above. 
Since we are looking for a particular geodesic which will minimise the distance between two points 
$\{\Omega_1,\xi_1\}$ and $\{\Omega_2,\xi_2\}$ the constant $\mathcal{M}$ 
can be expressed as a function of these coordinate values.

From Eq. (\ref{separate}) we have the following two separated equations
\begin{equation}
\Bigg(\frac{d S_{x_{1}}}{d x_{1}}\Bigg)^{2}=\frac{3}{32}\Bigg[\frac{K^{2}}{\left(1+2x_1\right)^{2}}-
\frac{\mathcal{M}}{x_{1}^{2}}\Bigg]~,~~\Bigg(\frac{dS_{x_{2}}}{d x_{2}}\Bigg)^{2}=\frac{\mathcal{M}}{8x_{2}^{2}}~~.
\end{equation}
First order differential equations for coordinates themselves can be readily obtained by using the formula 
\begin{equation}
g_{ab}\frac{dx^{b}}{d\tau}=\frac{\partial  S(x^{a},\tau)}{\partial x^{a}}~~,
\end{equation}
and these, when simplified after some algebraic steps  are given by
\begin{equation}
\label{1storder}
\begin{split}
&\frac{d x_1}{d \tau}=\sqrt{\frac{32}{3}}\Bigg(K^{2}\left(1+2x_1\right)^{2}
-\frac{\mathcal{M}\left(1+2x_1\right)^{4}}{x_{1}^{2}}\Bigg)^{1/2}~,\\
&\frac{d x_2}{d \tau}=\frac{2\sqrt{2\mathcal{M}}x_{2}}{x_{1}^{2}}\left(1+2x_1\right)^{2}~.
\end{split}
\end{equation}

To understand the significance of the separation constant we notice that the QMT admits a
Killing vector of  the form $K_a=(0,\frac{C_3x_1^2}{x_2(1+2x_1)^2})$, $C_3$ being a constant.
The conserved quantity along the geodesic  associated this Killing vector is
$\frac{C_3x_1^2}{x_2(1+2x_1)^2} \frac{dx_2}{d\tau}$, which as can be seen from the second relation 
of Eq. (\ref{1storder}) is just proportional to $\sqrt{\mathcal{M}}$. Hence the conserved quantity associated with the Killing vector
is just the separation constant associated with the HJ equations which corresponds to 
a hidden symmetry of the metric tensor. In fact, the existence of  the Killing vector implies that we 
can transform to a new set of coordinates, say $(X_1,X_2)$ from the present one, $(x_1,x_2)$
such that one of the coordinates of the new set becomes cyclic thereby making the 
symmetry explicit.

Since the general solutions of this set of equations are difficult to obtain for generic values of 
$\mathcal{M}$, we shall solve these equations for a particular value of the separation constant, namely 
for simplicity we choose $\mathcal{M}=0$. However, this is not a significant restriction, the reason being 
the following. As we have mentioned before, after we solve the first order equations, the two new integration 
constants present in the solutions $x_{1}(\tau),x_{2}(\tau)$  (call them $C_{1}$ and $C_{2}$ for convenience), 
as well as two constants $K$ and $\mathcal{M}$ already present are determined in terms of the 
coordinates $\{\Omega_1,\xi_1\}$ and $\{\Omega_2,\xi_2\}$. Alternatively, we can choose one of constants of 
the first set, say $\mathcal{M}$ to be given beforehand (note that since in the end the constant $K$ 
is the required complexity which we want to determine, we can not take this to be given).
so that one from the other set (we take $\Omega_{2}$) is determined in terms of the others, 
here it is the set $\mathcal{M},\Omega_{1},\xi_1,~\text{and}~\xi_2$.

To calculate the FSC, we need the solutions
 of Eqs.  (\ref{1storder}) with $\mathcal{M} =0$ 
 (here $\mathcal{M}$ is the separation constant used in the Hamilton-Jacobi equation). These are given by  
\begin{equation}\label{n=0geodesics}
x_{1}(\tau)=C_{1}\exp\Bigg\{8\sqrt{\frac{2}{3}}K\tau\Bigg\}-\frac{1}{2}~,~x_{2}(\tau)=C_{2}=x_{2}(0)~.
\end{equation}
Here, $C_1$ and $C_2$ are two integration constants, $K=\sqrt{g_{ab}\dot{x}^{a}\dot{x}^{b}}$ is a constant along the geodesic,
and integrated between reference state with $\tau=0$ and target state with $\tau=1$ this is the geodesic distance.
The coordinate $x_{2}$ remains constant along the geodesic which we call $x_{2}(0)$ - the value at $\tau=0$ (the constant $C_{2}\neq0$). Since we have separated the geodesic equations in the transformed $\{x_{1}(\tau),x_{2}(\tau)\}$ coordinates not in the original  $\{\Omega,\xi\}$ coordinates the solution $x_{2}=C_{2}$ does not necessarily mean that any of the two coordinates in the original parameter manifold is a constant.

Having obtained the solutions the last step of finding the complexity is to use the boundary conditions at the end points to obtain $K$,
which itself is the geodesic distance and hence the complexity.
At the starting point of the geodesic $\tau=0$ we have 
\begin{equation}
\Omega_{1}=\Bigg[\frac{C_{2}^{2}}{C_{1}-\frac{1}{2}}-1\Bigg]^{1/2}~,~~\xi_1=C_{2}\left(C_{1}-\frac{1}{2}\right)^{1/2}~~.
\end{equation}
Inverting them we have the unknown constants in terms of the initial  coordinate locations 
\begin{equation}\label{constants}
C_{1}=\frac{\xi_1}{\sqrt{\Omega_{1}^{2}+1}}+\frac{1}{2}~,~~C_{2}=\left(\Omega_{1}^{2}+1\right)^{1/4}\sqrt{\xi_1}~~.
\end{equation}
These constants can also be directly obtained from the relations given in Eq. (\ref{inversetra}).

On the other hand at the end point $\tau=1$ we have 
\begin{equation}
\xi_2=C_{2}\Bigg[C_{1}\exp\bigg\{8\sqrt{\frac{2}{3}}K\bigg\}-\frac{1}{2}\Bigg]^{1/2}~.
\end{equation}
Inverting this relation we obtain the complexity $\mathcal{C}_{FS}=K$ as
\begin{equation}
\label{ComFSMain}
\begin{split}
\mathcal{C}_{FS}&=\sqrt{\frac{3}{2}}\frac{1}{8}\ln\bigg[\frac{1}{C_{1}}
\bigg\{\bigg(\frac{\xi_2}{C_{2}}\bigg)^{2}+\frac{1}{2}\bigg\}\bigg]\\
&=\sqrt{\frac{3}{2}}\frac{1}{8}\ln\Bigg[\frac{1}{\xi_{1}}\Bigg\{\frac{2\xi_2^{2}+
\xi_1\sqrt{\Omega_{1}^{2}+1}}{2\xi_1+\sqrt{\Omega_{1}^{2}+1}}\Bigg\}\Bigg].
\end{split}
\end{equation}
Note that with the assumptions we have obtained the complexity (i.e., choosing the constant 
$\mathcal{M}$ to be independent) if the parameter $\xi$ is the same at the starting and the ending points i.e., 
$\xi_1=\xi_2$ then the complexity vanishes as can be directly seen by putting $\xi_1=\xi_2$ in Eq. (\ref{ComFSMain}). 

\noindent
\textbf{Separation of geodesic equations for  $\xi<0$:}
Here the relations between the \{$\tilde{x}_1,\tilde{x}_2\}$ and \{$\xi,\Omega$\} coordinates are 
given by
\begin{equation}\label{x12tilde}
	\tilde{x}_1=\frac{-\xi}{\sqrt{\Omega^{2}+1}}~, ~ \tilde{x}_2=(\Omega^{2}+1)^{1/4}\sqrt{-\xi}~,
\end{equation}
And the inverse relations are given by 
\begin{equation}
\label{inversetra2}
	\xi=-\sqrt{\tilde{x}_1}\tilde{x}_2~, ~\Omega=\sqrt{\frac{\tilde{x}_2^2}{\tilde{x}_1}-1}~.
\end{equation}
In terms of the $\tilde{x}_1$ coordinate the separatrix is at $\tilde{x}_{1}=1/2$.  

An entirely similar procedure as that of the previous case can be followed to separate the
geodesic equations for the metric in Eq. (\ref{line4}) as well. 
The separated equations analogous to those of  Eqs. (\ref{1storder}) are given by
\begin{equation}
\label{1storderxi<0}
\begin{split}
&\frac{d \tilde{x}_1}{d \tau}=\sqrt{\frac{32}{3}}\Bigg(K^{2}\left(1-2\tilde{x}_1\right)^{2}
-\frac{\mathcal{M}\left(1-2\tilde{x}_1\right)^{4}}{\tilde{x}_{1}^{2}}\Bigg)^{1/2}~,\\
&\frac{d \tilde{x}_2}{d \tau}=\frac{2\sqrt{2\mathcal{M}}\tilde{x}_{2}}{\tilde{x}_{1}^{2}}\left(1-2\tilde{x}_1\right)^{2}~.
\end{split}
\end{equation}
 Solutions of Eq. (\ref{1storderxi<0}) with $\mathcal{M}=0$ are given by
 \begin{equation}\label{n=0geodesics2}
 	\tilde{x}_{1}(\tau)=\tilde{C}_{1}\exp\Bigg\{-8\sqrt{\frac{2}{3}}K\tau\Bigg\}+\frac{1}{2}~,
 	~\tilde{x}_{2}(\tau)=\tilde{C}_{2}=\tilde{x}_{2}(0)~.
 \end{equation}
 From Eq. (\ref{x12tilde}) the conditions at $\tau=0$ give the two unknown constants in terms of the initial coordinate locations 
 \begin{equation}\label{constants-tilde}
 	\tilde{C}_{1}=\frac{-\xi_1}{\sqrt{\Omega_{1}^{2}+1}}-\frac{1}{2}~,
 	~\tilde{C}_{2}=\left(\Omega_{1}^{2}+1\right)^{1/4}\sqrt{-\xi_1}~~.
 \end{equation}
 Now evaluating the solution of Eq. (\ref{n=0geodesics2}) at $\tau=1$ and inverting we have the expression
 for the geodesic distance and hence the FSC quoted in the text.
 
\section{FSC of LMG exited state}
\label{LMG2app}
We now briefly discuss the calculation of FSC of the highest energy state of the LMG model in Eq. (\ref{LMG2}). 
We shall concentrate only on the symmetric phase since the analytical expressions of the QMT is difficult to obtain for 
the broken symmetry phase  \cite{GGCHV}. It has been shown in Ref. \cite{GGCHV} the exited state shows a QPT at 
$\xi=\frac{1}{2}\sqrt{\Omega^{2}+1}$ and our aim in this appendix will be to see the behaviour of the FSC close to 
this transition.

The computation of the QMT goes exactly in the same fashion as in the ground state. Here the line element  
analogous to that in Eq. (\ref{line3}) after the diagonalisation is performed is given by
\begin{equation}
\label{line3ext}
d\tau^{2}_{ext}=\frac{1}{32x_2^{2}\left(1-2x_1\right)^{2}}\Big[3x_2^{2}\text{d}x_1^{2}+4x_1^{2}\text{d}x_2^{2}\Big]~,
\end{equation}
where the subscript $ext$ refers to the exited phase and the coordinates $\{x_1,x_2\}$ are related to the 
set $\{\Omega,\xi\}$ with the same formula as those  one given in  Eq. (\ref{inversetra}). Once again the condition 
$x_{1}\neq0$ is valid. The Ricci scalar computed from this metric, given by $R=-\frac{128}{3x_{1}}$, is regular across 
the parameter manifold of the exited phase and thus it fails to notify the QPT.

The resulting geodesic equations can be once again separated using the HJ method and the solutions 
are now given by (compare with those of  Eq. (\ref{n=0geodesics}) for the ground state)
\begin{equation}
x_{1}(\tau)=C_{1}\exp\Bigg\{-8\sqrt{\frac{2}{3}}K\tau\Bigg\}+\frac{1}{2}~,~x_{2}(\tau)=C_{2}=x_{2}(0)~.
\end{equation}

The complexity in terms of the boundary point coordinates is now given by
\begin{equation}
\label{n=1comFS}
\begin{split}
\mathcal{C}_{FS}&=-\sqrt{\frac{3}{2}}\frac{1}{8}\ln\bigg[\frac{1}{C_{1}}\bigg\{\bigg(\frac{\xi_2}{C_{2}}\bigg)^{2}
-\frac{1}{2}\bigg\}\bigg]\\
&=\sqrt{\frac{3}{2}}\frac{1}{8}\ln\Bigg[\xi_{1}\Bigg\{\frac{2\xi_1-
\sqrt{\Omega_{1}^{2}+1}}{2\xi_2^{2}-\xi_1\sqrt{\Omega_{1}^{2}+1}}\Bigg\}\Bigg]~.
\end{split}
\end{equation}
We see that as the reference state point $\{\Omega_1,\xi_1\}$ moves towards the separatrix which is in this case, the
line $\xi=\frac{1}{2}\sqrt{\Omega^{2}+1}$, the expression for the FSC diverges, and it has the same features that
we have seen in Section \ref{sec4}.


\end{document}